\documentclass[acmsmall, nonacm]{acmart}
\AtBeginDocument{%
  }

\setcopyright{acmlicensed}
\copyrightyear{2025}
\acmYear{2025}
\acmDOI{XXXXXXX.XXXXXXX}

\setcopyright{none} 

\renewcommand\footnotetextcopyrightpermission[1]{} 

\begin{document}

\title{Report on the Designing Accountable Software Systems Workshop}

\author{Catherine Albiston}
\authornote{Authors listed in alphabetical order.}
\affiliation{%
  \institution{UC Berkeley}
  \country{USA}
}
\email{calbiston@berkeley.edu}

\author{Travis Breaux}
\affiliation{%
  \institution{Carnegie Mellon University}
  \country{USA}
}
\email{breaux@cs.cmu.edu}

\author{Kat Dearstyne}
\affiliation{%
  \institution{University of Notre Dame}
  \country{USA}
}
\email{kdearsty@nd.edu}

\author{Jane Cleland-Huang}
\affiliation{%
 \institution{University of Notre Dame}
  \country{USA}
}
\email{janeclelandhuang@nd.edu}

\author{Serge Egelman}
\affiliation{%
  \institution{UC Berkeley / ICSI}
  \country{USA}
}
\email{egelman@icsi.berkeley.edu}

\author{Joan Feigenbaum}
\affiliation{%
  \institution{Yale University}
  \country{USA}
}
\email{joan.feigenbaum@yale.edu}

\author{Lu Feng}
\affiliation{%
  \institution{University of Virginia}
  \country{USA}
}
\email{lf9u@virginia.edu}

\author{Max Lindquist}
\affiliation{%
  \institution{UC Berkeley}
  \country{USA}
}
\email{max.lindquist@berkeley.edu}

\author{Stephen Miner}
\affiliation{%
  \institution{Yale University}
  \country{USA}
}
\email{stephen.miner@yale.edu}

\author{Ruzica Piskac}
\affiliation{%
  \institution{Yale University}
  \country{USA}
}
\email{ruzica.piskac@yale.edu}

\author{Sarah Santos}
\affiliation{%
  \institution{Carnegie Mellon University}
  \country{USA}
}
\email{ssantos@andrew.cmu.edu}

\author{Jordan Schmerge}
\affiliation{%
  \institution{Yale University}
  \country{USA}
}
\email{jordan.schmerge@yale.edu}

\author{Anmol Singhal}
\affiliation{%
  \institution{Carnegie Mellon University}
  \country{USA}
}
\email{singhal2@andrew.cmu.edu}

\author{Maria Smith}
\affiliation{%
  \institution{UC Berkeley}
  \country{USA}
}
\email{mariasmith@berkeley.edu}

\author{Daniel Weitzner}
\affiliation{%
  \institution{MIT}
  \country{USA}
}
\email{djweitzner@csail.mit.edu}

\author{Christopher Yoo}
\affiliation{%
  \institution{University of Pennsylvania}
  \country{USA}
}
\email{csyoo@law.upenn.edu}

\renewcommand{\shortauthors}{Albiston et al.}

\begin{abstract}
  The Workshop on Designing Accountable Software Systems (DASS) was convened in November 2024 with support from the U.S. National Science Foundation to engage a wide range of current and future stakeholders from government, academia, and industry on the cross-disciplinary topic of accountability in software systems. Over two days, attendees engaged in a series of panels, invited talks, and breakout sessions covering: (1) the dimensions of accountability, including legal compliance as well as business and societal aspects and drivers; (2) a conceptual model of the various structures needed to realize accountability; (3) the sources of legal requirements that affect software; (4) the operationalization of legal requirements in software; (5) the requirements to preserve evidence needed to conduct investigations; and (6) a range of challenges and contextual factors beyond software that affect why some accountability structures succeed, while others fail. The workshop was conducted as a collaborative systematization of knowledge that culminated in several research directions. The findings include the importance of clarifying definitions and responsibilities within accountable organizations, which can affect whether those researching accountability are making assumptions that limit the generalizability of findings. Further research was also identified as needed to study the ways to improve the translation of accountability structures into the software design process while improving engagement with stakeholders, such as legislators, regulators, business executives and system developers. Finally, a key finding was the high demands that DASS-like research projects place on interdisciplinary teams: both in terms of team formation and sustainment, as well as, the specific demands of cross-disciplinary learning that covers both research methods, research dissemination, and career development.
\end{abstract}





\maketitle

\section{Introduction}
Businesses, government and the public have come to depend on software applications and systems to conduct themselves in nearly every aspect of society and daily life. With this pervasive dependence on software, comes the need for software to satisfy the requirements of laws, regulations, and norms so that software behavior is consistent with society's expectations. Because laws are written to govern organizations and individuals and because norms may be unwritten, despite being generally understood, the challenge of designing software that is accountable to systems of governance is not always straightforward. To investigate this challenge, the National Science Foundation (NSF) introduced the Designing Accountable Software Systems (DASS) program beginning in 2021 (see solicitation NSF 22-512 and NSF 21-554). The DASS program was a multi-directorate program, operating from the Directorate for Computer and Information Science and Engineering (CISE) and the Directorate for Social and Behavioral Economic Sciences (SBE). Between 2021-2023, the NSF DASS program awarded 42 awards across 34 research institutions. In this report, we describe the findings of a workshop aimed at reviewing the history of the DASS program. Whereas traditional systematization of knowledge can arise from a thorough and comprehensive literature review, the workshop consisted of an interactive systematization, directed at reviewing participant experiences of conducting research in accountability as well as enumerating open research questions and hard challenges to be addressed going forward. Unlike the written record, in which open research challenges are only briefly mentioned or are inferred through a lack of attention in the record, a workshop with impaneled speakers and small group working sessions can surface issues that researchers struggle to solve for which no written record yet exists.

On November 17-19, 2024 the Workshop on Designing Accountable Software Systems was held in Warrenton, Virginia to engage stakeholders in and around the research community to understand current and future DASS challenges. With three years of the NSF DASS program underway, new evidence clarifying the research challenges under DASS as well as evidence of recent advances had accrued and could now be brought into focus to guide future programming. The workshop program was assembled in summer 2024 by a steering committee of nine members who are among the co-authors of this report and who hold expertise that draws from across computer science, law, and sociology and from across eight universities. Workshop invitations went to current and former recipients of NSF DASS awards, in addition to other government, industry, and academic members with current or prospective interest in the topic. In total, 78 participants attended.
Prior to the workshop, the steering committee surveyed current and former NSF DASS program principal investigators (PIs) about their experiences and findings, which are reported, herein. Through an analysis of the survey results, we observed that the PIs cited several benefits of their interdisciplinary collaborations. PIs from computer science (CS) appreciated the opportunity to work with and observe the perspective of lawyers, with whom they seldom collaborate. More generally, bringing several disciplinary perspectives to the table reportedly enriched projects and led to new ideas and research directions. The PIs noted that they ``learned to look at problems from a non-CS perspective,'' they ``broadened [their] visions,'' they asked ``more ambitious questions'' and they learned ``other facets of [the] problem'' that they presumably would not have encountered otherwise. One PI noted that ``[w]e are able to leverage disciplinary knowledge and professional connections in both computer science and law'' to expand the reach and impact of the research.

Challenges for PIs mostly centered around different publication models and expectations in different disciplines. For example, one PI noted that ``one challenge is the differences in venues for publications. For social scientists, conference papers are not significant, while CS researchers aim to publish in top-tier conferences.'' The PIs also mentioned that disciplines often use the same words to mean different things, so understanding each other took some time. The PIs said that experience with interdisciplinary collaborations or previously working together tended to minimize the start-up costs and other challenges related to collaboration.
    
\begin{figure}[h]
    \centering
    \includegraphics[width=0.6\textwidth]{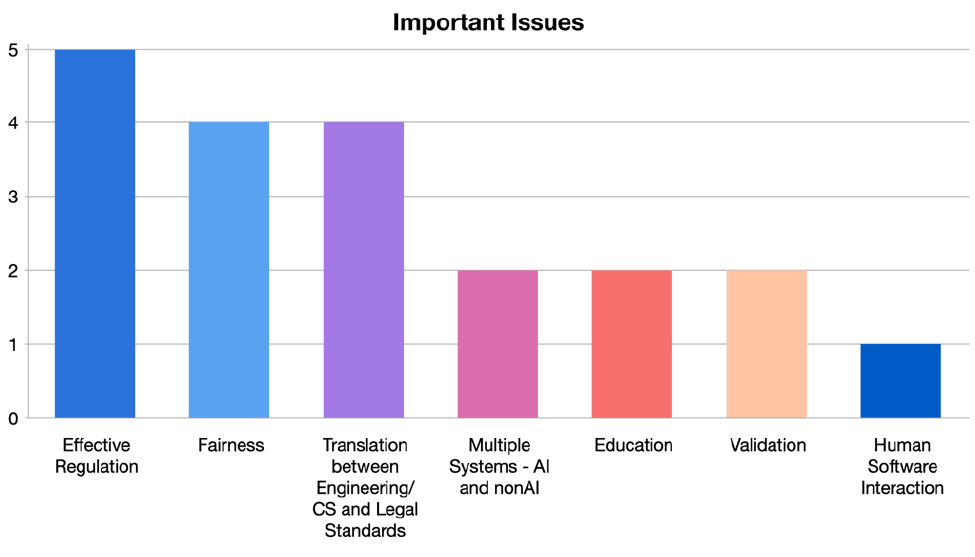}
    \caption{Issues Identified by Surveyed NSF DASS Principal Investigators}
    \label{fig:diagram1}
\end{figure}

Figure\ref{fig:diagram1} summarizes the categories of what PIs reported as the most important challenges or issues at the intersection of software development, legal compliance, and social values. The y-axis represents the number of mentions of an issue across all survey responses. These issues correspond to concerns raised in the broader workshop, and represent concerns ranging from the technical (validation, translation) to the normative (fairness).

The workshop program yielded discussions over a number of topics, including the dimensions and structures underpinning accountability, the sources of legal requirements, the challenges of operationalizing legal requirements through software and system design, the manner of evidence needed to trust that a system is accountable, the consequences and people or organizations responsible, as well as, the manner in which accountability does and does not work in some areas. In the end, the meeting concluded with a diverse list of open research challenges, which we include at the end of the report. We now review the format and findings from each workshop session. The findings were obtained by two observational note-takers who were assigned to each session. The notes were later organized and summarized using the affinity diagramming method. After the notes were compiled into prose, the authors supplemented the prose with supporting citations to relevant background scholarship.

\section{Dimensions of Accountability}
The first session explored dimensions of accountability that extend beyond compliance with law. Presentations emphasized that accountability is a system property that can be framed either as an ability or a right. It also arises across a temporal spectrum that includes prevention, violation, detection, evidence, judgment, and punishment. Responsibility for accountability is not simply a matter for lawyers to implement. For the business community, accountability can mean more about protecting business reputations than liability. Moreover, for businesses operating in a vertical stack, accountability can also include expectations about operational issues such as disclosures, responsibility for handoffs, update schedules, and error response times. Technologies can be designed to support real-time operational decisions as well as tamper-resistant evidence of what occurred after the fact. In addition, societal norms and contextual expectations can determine expectations about accountability as well as law. Technological changes can violate existing norms in ways that accountability can address, and they can also enable new institutions that create a broader range of socially legitimate outcomes. The technical community thus plays a key role in designing systems that support accountability. Success of these efforts depends on research exploring the relationship between software and different visions of accountability and designing technologies to meet the needs of these different visions.

\section{Structures of Accountability}
In the second session, we examined the structures of accountability in order to understand what is in the design space of accountable systems and what larger dynamics may affect these designs. By understanding these structures, we aim to clarify what kind of evidence must be collected in order to hold a system operator or provider to account in the eyes of an authority for adherence to a rule or behavior. To understand the structures of accountability, we examined: 1) sources of rules and behavior for which we seek accountability; 2) gaps that may emerge in the definition and development of those sources; and 3) particular challenges associated with the trustworthiness of accountability structures in society. With these structures in view, we discuss what would be the ideal characteristics of an accountable systems laboratory.

Sources of accountability begin with laws and other formally codified rulesets, but that is only the beginning of the kinds of behavioral expectations that accountable system design must accommodate. Industry self-regulatory bodies create a variety of guidelines, best practices, and recommendations which to various entities hold themselves out as being accountable. Professions such as medicine and engineering (including many fields such as mechanical and civil, but not including software) have professional standards and large bodies of knowledge according to which practitioners are certified and credentialed. This heterogeneous body of behavioral expectations is used as measures against which both individual professionals and organizations are judged. Add to this body, technical standards on safety, security, and performance measures, and interoperability rules, and you have a larger body of rules and guidelines defining behaviors that are the subject of accountability. In order to design accountable systems, we need a more comprehensive understanding of the range of behaviors that are the subject of accountability and a clearer view of how to generate the right kind of evidence to preserve behavior relevant to an assessment.

These sources of accountability that exist may not, however, always provide clear ground for assessing fidelity with the rule sets or for disciplining the behavior that matters most to society. In discussing the institutions that are the sources of accountability, we encountered gaps between the ideal rules and those rules that are actually followed in practice. In some cases, these gaps are the result of divergent evolution of the rules written `in the books' versus those rules or norms that develop `on the ground,' as Bamberger and Mulligan have explained.\footnote{Bamberger, Kenneth A., and Deirdre K. Mulligan. \textit{Privacy on the ground: driving corporate behavior in the United States and Europe.} MIT Press, 2015.}  In other cases, private order emerges without regard to or in ignorance of laws on the books, as Ellickson showed.\footnote{Ellickson, Robert C. \textit{Order without law: How neighbors settle disputes.} Harvard University Press, 1991.}  Some of the divergence between legal rules and the institutions that emerge to express their behavior is more calculated. Tendency toward legal endogeneity, emergence of private sector compliance structures that have the effect of subverting the rules enacted in law, can blunt the large intent of accountable systems by creating the appearance of rule following without actually serving the underlying goal of those rules.\footnote{Edelman, Lauren B. \textit{Working law: courts, corporations, and symbolic civil rights.} University of Chicago Press. 2016.}

In order to address gaps between idealized rules, on the one hand, and actual compliance and enforcement processes, on the other, research on the design of accountable systems should consider a wide range of market incentives and implementation costs. Discussion as part of this panel highlighted the significant gap between rules on the books and the ability of enforcement authorities to detect possible violations at scale in complex digital systems. This more holistic view of the context of accountability will help understand which system designs are actually likely to serve the interests of accountability, and how to achieve the goal most efficiently.

Finally, recognizing the complex landscape in which accountability systems must function, the panel considered what would be an ideal laboratory for the design and evaluation of new accountable systems approaches. Cross disciplinary teams with technical expertise in software engineering, applied cryptography and HCI, together with economists, legal scholars, and other social scientists will be needed to understand the interplay between technical design, human and organizational behavior, and the law. In order to test designs and understand institutional dynamics, the ideal lab would include participation of up to one dozen companies representing a single sector, policy makers and consumer advocates with a range of regulatory perspectives.

\section{Sources of Legal Requirements}

This session featured industry and government perspectives on software accountability: how industry tries to determine legal requirements for software (e.g., what privacy regulations to comply with), as well as how regulators think about legal requirements (including how they pursue cases—one type of accountability). We discussed prior research that examined software developers and found that many software compliance issues are due to developers being unaware of or confused about their compliance obligations. One reason why compliance issues exist is because they are seen as purely legal issues, rather than engineering problems. Any solution is going to involve proactively integrating compliance and privacy concerns at every stage of software development, and may also involve implementing data mapping-like tools for privacy concerns. We discussed open research questions related to compliance auditing and methodologies for studying compliance, how developers should and do learn about compliance requirements

Both industry and government have a need for better compliance tools, but no one really knows what those might look like. Currently, large companies that can afford to build privacy teams are solving this by engineering their own compliance tools; this is costly and thus smaller companies simply ignore compliance requirements. Regulators need tools for their own investigations (and do not develop them internally), but also need to understand industry best practices. Right now, a lot of organizations are doing ``symbolic compliance,'' where their practices do not necessarily match what they publicly claim or what shows up in their audit reports. Better tools to inventory data and to better explain compliance issues to various stakeholders would decrease the compliance burden and also better align stated practices with actual practices.
	Ultimately, accountability can be increased by making compliance tools more available. Developers need these tools to understand their compliance obligations and how to address them, and regulators need better tools to more easily identify non-compliant products and services. Research at the doctoral level is critical to uncovering privacy vulnerabilities, understanding impact of data collection and legal violations, and evaluating existing regulations and areas for improvement, particularly as it relates to societal values and human rights. Ultimately, as it has done with security, research sponsors should support research aimed at understanding how software compliance issues come to be, as well as research into data-driven software development frameworks that are designed to mitigate and/or minimize software compliance issues.

\section{Operationalizing Legal Requirements in Software}
Ensuring that software systems comply with legal requirements is not just a matter of implementing rules but requires a systematic process to translate regulations into technical design, governance, and continuous oversight. Furthermore, software systems do not operate in isolation; they interact with complex and evolving regulatory landscapes, creating challenges in maintaining compliance, safety, and accountability. A key element of this process is setting and managing expectations, including the expectations of regulators who establish compliance requirements, the public who must understand the system's capabilities and limitations, and system designers who must translate these expectations into concrete mechanisms that track compliance and performance. For example, self-driving car manufacturers have documented cases where their systems struggled to detect and respond to perpendicular traffic at intersections, but these limitations have often not been explicitly conveyed in a way that ensures drivers remain appropriately vigilant and understand when intervention may be necessary.

One of the most persistent challenges in operationalizing legal requirements is the evolving nature of software. Traditional safety certification methods, which rely on human inspection of code, struggle to keep pace with continuously evolving systems, especially AI-driven systems that do not follow deterministic rules. Potential solutions, such as Pre-Determined Change Control\footnote{USFDA, Health Canada and UK MHRA, ``Predetermined Change Control Plans for Machine Learning-Enabled Medical Devices: Guiding Principles,'' October 2023.} (PCCP), are emerging as structured ways to assess modifications and determine when recertification is necessary. However, general safety standards, such as ISO 26262\footnote{ISO 26262-1:2018 Road vehicles — Functional safety}, do not adequately address the dynamic nature of AI systems, nor the unpredictable ways in which software can fail due to mismatches between assumptions and real-world conditions.

Automated systems must also manage different types of uncertainty. In domains like payroll tax software, ambiguity stems from constantly evolving legal language and jurisdictional complexity. Herein, domain experts can help bridge the gap between legal requirements and system implementation. In contrast, AI-driven systems, such as self-driving cars, introduce a deeper challenge of indeterminate behavior that engineers cannot always predict or model probabilistically. This makes verification and validation particularly difficult. For instance, early versions of Waymo's self-driving software learned to stop for unknown obstacles, causing traffic disruptions. Engineers modified the behavior to pull over instead, but in one case, a vehicle stopped for a pedestrian, but then hit and dragged the pedestrian along the road as it attempted to pull over, prompting new regulatory action. This and similar examples illustrate the difficulty of ensuring that AI systems behave predictably in compliance with regulations in unbounded real-world environments.

Successfully operationalizing legal requirements in software therefore requires a shift in how we approach compliance, risk, and responsibility. Not only must systems be designed to remain robust when their underlying assumptions fail, but expectations must also be managed to ensure alignment between software capabilities and public understanding. Achieving the goal of true accountability requires interdisciplinary collaboration, leveraging expertise from engineers, policymakers, legal scholars, and social scientists to build systems that are not only compliant but also aligned with societal values. 
6.	Evidence Needs in Conducting Investigations

One of the core problems in ``Designing Accountable Software Systems'' is the need for a software system to produce, preserve, and present evidence that it has behaved properly when it is ``called to account.'' This section summarizes the workshop participants' thoughts on how to flesh out and, ultimately, to solve this core problem.

As always in discussions of accountability, one must ask ``to whom is the software system supposed to be accountable and for what?'' The best understood and most thoroughly researched case is that in which there is a law that specifies what the software must (or must not) do or what properties it must have. Desai and Kroll\footnote{Deven R. Desai and Joshua A. Kroll, ``Trust But Verify: A Guide to Algorithms and the Law,'' \textit{Harvard Journal of Law and Technology}, Vol. 31, No. 1, Winter 2018.} posit a division between technical accountability and legal accountability. Technical accountability addresses the need for software to produce evidence that allows oversight and verification that the software is operating according to the law in question. Such evidence must have integrity, i.e., must be a tamper-evident record that provides non-repudiable evidence of relevant actions by the system. What actions were taken and why? Technical accountability does not directly provide legal accountability but rather enables it by providing a way to understand whether, how, to what extent, and why misdeeds occurred, as well as who (or what part of a system) is responsible for them. It is then up to political and legal processes to use that evidence to hold actors responsible, meting out punishments when warranted.
Exactly what evidence an accountable software system should produce, preserve, and present to legal authorities varies widely across application domains. Important questions to bear in mind include but are not limited to: (1) What level of granularity is called for? Automation and digitization of processes that were once executed and monitored by humans have led to huge increases in evidence-gathering capacity, but ``just log everything'' won't work. Evidence must be relevant and understandable in order to be useful in achieving legal accountability. (2) Does the law in question address design accountability or runtime accountability? (3) Does the law in question put forth positive or negative requirements? That is, will the software be held accountable for doing something (such as displaying accurate information to workers who depend upon it) or for not doing something (such as rejecting all job applicants in a class that is protected under the Civil Rights Act)? Negative accountability, i.e., evidence of a prohibited behavior, is much more complicated and, in some applications, simply not fully automatable. Fortunately, there are many cases in which a tamper-evident record of certain actions taken by a software system can support (if not rigorously prove) a claim that it satisfies a negative requirement; the key point is that identification of those actions and formulation of the argument that they support the claim must be done carefully when the system is designed.

Sometimes a software system is found to be causing harms that were not previously foreseen and thus never precluded by an existing law. In such cases, those who produced and/or are maintaining the system must modify it so that it does not cause the harm in question and enable it to produce technical evidence that it is not doing so.

Finally, accountable software-system developers might have much to learn from technological domains in which accountability has worked well, e.g., ``black boxes'' in airplanes, nuclear power plants, and ``high-assurance'' or ``clean-room'' software engineering. End-to-end system design and implementation in those domains is too expensive to be widely applicable, but individual techniques, practices, and insights may be helpful to DASS investigators.

\section{Why do some things work and other things not in accountability?}
Several important themes emerged from this session, including definitional issues, measurement, implementation challenges in social context, ambiguity and the certainty of imperfection, concerns about validity, and the veneer of objectivity associated with algorithms. A summary of these themes and the discussion follows.

Participants pointed out that how to define automated decision making encompasses two very different concepts: (1) ``top down'' translation of preexisting policy into automated systems for implementation; and (2) ``bottom up'' development of policy by learning from data. For the first concept, failures arise from errors in translation into code, difficulties implementing ambiguous terms, problems arising from how algorithms limit human discretion, errors flowing from the dispersion of accountability for developing systems, and the inability to predict in advance all flaws or challenges for the system. For the second concept, failures arise from problems with data that reflect past human judgments including bias, nuance lost in the collapse of the range of factors that go into decision-making, the fact that machine learning does not ``learn'' or think like the human mind and therefore can make inferences that are nonsensical, and how the need for discrete, measurable outcomes limits the scope of what is important according to ML needs, not policy priorities.

Participants elaborated on how decisions about measurement often implicitly incorporated policy and normative judgments that then become objectified as part of the algorithm. Conceptualization of key concepts such as harm or fairness is deeply contested so that the algorithmic resolution requires transparency and participation by affected groups. The thing that is measured is what processes are prioritizing, even if that is not the intention. Organizational and social values become embedded in algorithms, and code is less flexible and interpretable than law; there is no ``code court,'' for example, to adjudicate whether a particular implementation is correct. The validity/invalidity of measurement can be a major source of harm when what one believes is being measured is not consistent with what is actually being measured; construct and content validity are seldom considered when deciding on measurement, which can produce arbitrary results. A focus on efficiency over validity and accountability may increase the possibility of failure. Choices of measurement, or ``actuarial practices,'' are not objective or neutral, and can depoliticize important policy questions that should be subject to democratic debate. In this way, measurement becomes a form of governance. The veneer of objectivity that adheres to algorithms can reduce accountability by obscuring these normative decisions and political judgments. Actuarial practices create a new point of leverage for the exercise of power and can stifle attempts to hold systems accountable. Participants also warned of measurement creep when measurements are co-opted and repurposed for ends for which they were not intended.

Participants emphasized that every system would have imperfections and some failures. There is also an inability to mechanize some processes and outcomes, and it is impossible to stamp out ambiguity. The key, then, is to embrace this imperfection without compromising social consequences. One strategy is to implement policy or develop policy at a lower level of abstraction to allow the operator to interpret what was happening and adjust, integrating human judgment over system-provided information. Participants also noted that expecting imperfection encouraged considering the consequences if the system did not work as planned, which helps mitigate risks.

Participants pointed out that we have an overly simplistic mental model of top-down legal regulation, but software is developed in the social context of organizations, engineering departments, and companies. Legal regulations are translated through the values and needs of these social settings which reduces accountability through loose coupling between organization signals of compliance and actions on the ground in software development. This well-documented organizational behavior insulates organizations from legal accountability without ensuring substantive compliance. As largely symbolic indicia of compliance become more common, other organizations adopt them to gain legitimacy. Soon the meaning of ``accountability'' becomes these symbolic actions, and the regulated entities themselves are defining compliance as these symbolic actions. It is important to watch for this process and look beyond symbols of legal compliance to ensure real accountability. In addition, algorithmic accountability includes not only how data are collected and used, but also how they are presented because interpretations of even low-level measurements are political and come to convey what is normative. Attention to how results are presented, e.g. the interface with human users, is needed for accountability to ensure results are properly interpreted. Power also contributes to how algorithms are developed and implemented, and what norms are institutionalized. Some groups are excluded from decisions that affect them, resulting in harms to individuals and communities that are overlooked, and outcomes that are neither transparent nor acceptable. Attention to representation and power are essential for accountability.

\section{Research Directions and Questions}
The workshop participants invested significant effort in reviewing each session's highlights and enumerating open research challenges, expressed as broad questions. In a disclaimer to the reader: the questions below have not been vetted for their novelty or technical relevance to the goals of NSF programs, including future DASS programs. Rather, they represent the questions of the workshop membership that they viewed as in need of greater theoretical and technical attention, whether that be through an exhaustive literature review or new fundamental or applied research.

The questions are organized under themes, including questions about the definition of accountability, who is fundamentally responsible for accountability, how software development occurs, how laws and other expressions of requirements are translated into software, how processes should engage with others outside of core development, and how enforcement should be realized. In addition, the workshop attendees reflected on two key challenges to conducting successful DASS research, which are presented at the end.

\textit{Definitions}: Accountable system design begins with answering, who or what are software systems accountable to (society, law, individuals, organizations, public or private sector entities?) Individual disciplines may establish their own profession-specific definitions and thus one question raised by attendees concerns, \textit{what other definitions of accountability exist beyond legal compliance, for example, in journalism, sociology, and economics, among others?} In journalism, trust in the media has been eroding, re-defining accountability.\footnote{Fengler, S. (2019, March 26). ``Accountability in Journalism.'' \textit{Oxford Research Encyclopedia of Communication.}} Accountability in journalism includes media self-regulation and self-control as well as other actors to embody ``any non-state means to make the media accountable to the public;''\footnote{Bertrand, C. (2000). \textit{Media ethics \& accountability systems.} New Brunswick, NJ; London, U.K.: Transaction.} ``media transparency'' refers to visibility into newsroom processes and actor roles in preserving trust between the media and the public;\footnote{Meier, K., \& Reimer, J. (2011). ``Transparenz im journalismus: Instrumente, konfliktpotentiale, wirkung.'' \textit{Publizistik}, 56(2),
133–155.} and ``co-regulation'' occurs when media law coincides with self-control bodies appointed by the media industry.\footnote{Puppis, M. (2007). ``Media governance as a horizontal extension of media regulation: The importance of self- and co-regulation.'' \textit{Communication: European Journal of Communication Research}, 32(3), 330–336.} These and other practices are surveyed in detail by Fengler, 2019. \textit{Who or what are software systems accountable to (society, law, individuals, organizations, public or private sector entities?)} For software systems that support media creation and distribution, such as news sites, blog and podcast platforms and social media, one consequence of this question is that software design processes must integrate with profession-relevant definitions of accountability that characterize the historical and practical relationship intended to preserve trust between the profession and the public. 

Uncertainty is a core concern when designing autonomous and self-adaptive systems that must account for uncertain future states, which raises the question, \textit{how can systems with inherent uncertainty meet established accountability requirements?} In self-adaptive systems, software is designed to modify its behavior at runtime, which can introduce uncertainty through various sources including oversimplifying assumptions, model drift, noise, and operating environmental changes, among others.\footnote{Esfahani, N., Malek, S. (2013). ``Uncertainty in Self-Adaptive Software Systems.'' In: de Lemos, R., Giese, H., Müller, H.A., Shaw, M. (eds) \textit{Software Engineering for Self-Adaptive Systems II. Lecture Notes in Computer Science,} vol 7475.} Definitions of accountability must recognize that uncertainty may be unknown until runtime, creating incentives to define accountability as retrospective reviews of system failure. \textit{How will definitions of accountability and the challenges of realizing accountability be transformed by recent advances in generative artificial intelligence?} In multimodal large language models (LLMs), accountability may be interpreted as ``alignment'' with human values\footnote{Bai, Yuntao, Saurav Kadavath, Sandipan Kundu, Amanda Askell, Jackson Kernion, Andy Jones, Anna Chen et al. ``Constitutional ai: Harmlessness from ai feedback.'' arXiv preprint arXiv:2212.08073 (2022).} or as law-following behavior.\footnote{Cullen O'Keefe et al., ``Law-Following AI: Designing AI Agents to Obey Human Laws,'' 94 \textit{Fordham L. Rev.} 57 (2025)} Benchmarks are frequently used to establish accountability in the form of a performance threshold, in which LLMs must correctly answer hundreds of questions on morality,\footnote{Ji, Jianchao, Yutong Chen, Mingyu Jin, Wujiang Xu, Wenyue Hua, and Yongfeng Zhang. ``Moralbench: Moral evaluation of llms.'' \textit{ACM SIGKDD Explorations Newsletter} 27, no. 1 (2025): 62-71.} ethics\footnote{Jiao, J., Afroogh, S., Murali, A. et al. ``LLM ethics benchmark: a three-dimensional assessment system for evaluating moral reasoning in large language models.'' \textit{Sci Rep} 15, 34642 (2025).} and law.\footnote{Guha, Neel et al. (2023) ``LEGALBENCH: a collaboratively built benchmark for measuring legal reasoning in large language models.'' In \textit{37th International Conference on Neural Information Processing Systems (NIPS '23)}, Article 1915, 44123–44279.} Benchmarks are limited by their incompleteness and societal and cultural bias. Lastly, contestability research, which argues that the public should be able to use the same processes by which they contest human decisions (e.g., complaints and legal procedures) to also contest algorithmic decisions,\footnote{Henrietta Lyons, Eduardo Velloso, and Tim Miller. 2021. ``Conceptualising Contestability: Perspectives on Contesting Algorithmic Decisions.''\textit{ In ACM Human Computer Interaction 5, CSCW1}, Article 106 (April 2021), 25 pages.} raises the question, \textit{how does contestability support accountability?}

\textit{Responsibility}: What governance structures and personnel roles should exist in addition to regulatory frameworks within organizations and outside in the public and private sector? How to avoid a vacuum of responsibility? How to deal with contested concepts? Do organizations need special roles assigned to be advocates? How to align accountability with management and how to ``roll up'' practices into corporate-level assurances? What communication modes, workflows and tools should exist to enable collaboration and a common vocabulary between software developers and legal professionals? How do we prevent misapplied or misunderstood legal principles in design? How do legal professionals, engineers, users, etc. perceive risks similarly or dissimilarly? What must engineers learn about societal values to design accountable systems? How does certification meet the needs of establishing accountability? What role should ethics play in defining the engineer's responsibility to design systems accountable to society? What role do global governance models and structures play? Which laws or legal and regulatory mechanisms help or hinder accountability in software?

\textit{Development}: Where within the software development lifecycle (Agile, Waterfall, etc.) should accountability be introduced? How to define the role of others, who are not programmers, about their role in accountability, e.g., UI/UX designers and what they need to do? How to introduce accountability into design as early as possible? To what extent does accountability require extensions to or make use of our current knowledge about harms and hazard analysis? Emphasis in legal accountability is placed on ``foreseeability'' and the extent to which organizations should be aware that potential harms exist. How can we communicate with developers when their designs for accountability are effective, e.g., by engaging those affected or by collecting other kinds of runtime information? How to design software to be resilient and adaptive to changes in laws and norms? How to establish ``living traceability'' from design to implementation, e.g., to detect design changes from well-intentioned developers after designs are subject to implementation constraints? What are good models for supporting long-term maintenance of accountable systems?

\textit{Translation}: How should laws and regulations be translated into software design? How to distinguish between ``the law on the books'' and ``the law in action'' when deciding what to design? What are the design principles for transparency that can protect administrative agencies' discretion? Can design principles include opportunities for refusal? Do mathematical frameworks exist to yield a rigorous understanding of the components of accountability? If so, then we need paradigms for translating such frameworks into practice. How to connect developers to concept contestation and resolution mechanisms to inform clarification around implementation challenges? How can we ensure accountability across the supply chain?
How to evaluate whether a design satisfies or violates a law? What evidence is needed, both prospective evidence (e.g., design artifacts) and retrospective evidence (e.g., logging and telemetry). What metrics should we use to measure that a system is accountable? Which stakeholders should be present when deciding accountability, e.g., how to decide when this should include a lawyer, sociologist, psychologist, political scientist, computer scientist, etc.?

\textit{Engagement}: How to engage stakeholders, including those affected by the technology, to share in decisions about accountable designs, e.g., through well-structured co-design? How does society come to understand what is occurring within software in order to hold it to account? How do we incentivize organizations to share information about compliance, and how do we discourage deception? What is the appropriate role of the public and private sectors as they share the act of establishing accountability? What opportunities for consumer literacy and engagement exist and, to what extent do definitions of accountability assume that consumers are active participants in the market by rewarding companies who act responsibly?

\textit{Enforcement}: How effective are regulatory mechanisms, and are they adequate for addressing or redressing the harms that arise from accountability failures? Given the lack of understanding, what kinds of enforcement mechanisms can exist that replace ``blame and sanction'' with ``dialogue and assistance?'' What kinds of information should flow back from runtime systems to inform better enforcement-through-design? How can procurement processes be improved to prioritize accountability in systems? What are the individual and organizational incentives to encourage self-enforcement? How do power differentials affect or yield selective or non-existent enforcement? How does the social system enforce the desired behavior of software and what are the limitations of enforcement, given software does not pervade all aspects of social systems?

\textit{Other Concerns}. A key finding expressed by attendees was the extent to which DASS faces all of the challenges of conducting interdisciplinary research, including understanding the incentives that motivate researchers across domains, how expertise is learned and applied by different professions, how researchers across domains communicate research findings, and so on. This challenge incurs an overhead within a DASS project by requiring time to align vocabulary and develop joint research goals and methodology. In addition, a key finding was that research into accountable software systems often and necessarily involves research into legal, sociological and economic questions, among other disciplinary questions. Sponsoring agencies should be open to contributions from these fields as a part of a holistic and comprehensive research proposal.

\section{Conclusion}
The outcomes of the Workshop on Designing Accountable Software Systems (DASS) indicate a rich scientific frontier with open, unanswered research questions that cut across the social, legal and computer sciences. While the dimensions and structures of accountability appear to be broad, robust and resilient to change, the continuing evolution of technology introduces new challenges not previously addressed. For example, with increased adoption of artificial intelligence (AI), legislators, businesses and developers will need increased awareness of emerging accountability gaps that automation creates, whereas the structures of evidence and enforcement may need to be operationalized in new ways to account for new technology nomenclature created in the science and engineering of AI. Finally, the workshop outcomes represent and support the need for investment in cross-disciplinary research, as accountability is understood differently in sociology, law, and computer science, among others. Without cross-disciplinary engagement, the depth of technical understanding acquired by any one discipline risks losing efficacy when positioned against the challenges known well to another discipline that is more adept at studying issues in their respective area.

\section{Acknowledgments} The 2024 DASS Workshop organization, program and attendance was supported by the U.S. National Science Foundation Award \#2435539.

\end{document}